\documentclass[11pt]{article}
\usepackage{graphicx}
\begin{document}
\def\p{\partial}
\def\f{\frac}
\def\l{\left}
\def\r{\right}
\def\({\l(}
\def\){\r)}
\def\[{\l \lbrack}
\def\]{\r \rbrack}

\baselineskip 22pt
\begin{center}
{\Large \bf Surface plasmon resonance under conditions of electromagnetically induced transparency
} \vspace{1.0cm}

\renewcommand{\thefootnote}{$\dag$}

 Chunguang Du  \vspace{0.5cm}

{\small \it  State Key Laboratory of Low-Dimensional Quantum Physics, Department of Physics, Tsinghua University,Beijing 100084, P. R. China.}
\end{center}

\begin{flushleft}
A scheme for a surface plasmon resonance system under conditions of electromagnetically induced transparency (EIT) is proposed. The system is composed of three layers: a prism, a thin metal film,
 and a hybrid dielectric consisting of EIT atoms and a background substance.  A probe and a coupling laser beam are input.  Corresponding analytical formulas are derived for the cases when one or both of the laser beams excite surface plasmon polaritons at the metal/dielectric interface. Under resonance conditions, an extremely sharp dip appears in the reflectivity-frequency spectrum of the probe field, revealing new properties of two-dimensional EIT. The reflectivity is extremely sensitive to shifts in the laser frequencies and atomic levels, and to variations of permittivity of the substrate. This EIT-SPR system may to be used for novel magnetometers and biosensors.


\end{flushleft}

Electromagnetically induced transparency (EIT) is a fascinating effect in which an otherwise opaque medium becomes transparent to
a resonant (probe) field by use of another (coupling) field \cite{EITreview}. The destructive quantum interference between two atomic transition pathways leads to null absorption and large dispersion within the induced transparency window.
The phenomenon has attracted great attention on account of its potential applications in the coherent control of the optical properties of atomic media and solid systems, e.g. photonic crystals doped with EIT atoms \cite{PCEIT}, or quantum metamaterials \cite{PRBdu}.When a weak probe laser propagates through a dilute EIT medium (EITM)
 the transmission spectrum has a high contrast ultra-narrow peak, which has also attracted great attention \cite{EITreview}. In contrast, its reflectivity spectrum has been seldom investigated, a key problem being that in ordinary cases the reflectivity is insensitive to the coupling field because the relative refractivity of the medium can only be modulated within a very small range (less than $6\times10^{-3}$ for cold-atoms \cite{slow}).
However, it will be shown in this letter that the situation is completely changed when the EIT medium is covered by a metal film, where surface plasmon (SP) polaritons can be excited under resonance conditions (i.e. through the well-known surface plasmon resonance (SPR)\cite{Raether}). These SP polaritons
are highly confined, non-radiative electromagnetic excitations associated with electron charge density waves propagating along a dielectric-metal interface.
  A remarkable feature is their highly sensitive response to changes in the refractive index near the metal surface, which has already been exploited in, for example, biosensors \cite{sensor}.
   In this letter, I will investigate a three-layer structure composed of prism/metal film/EITM, and show that even an EIT material of very low atomic number density can strongly modify the reflectivity and lead to an ultra-narrow dip in the frequency spectrum of the reflectivity. Also, it will be shown that the spectrum is extremely sensitive to the permittivity of the dielectric layer.
   To my knowledge, an SPR system based on EIT has never been investigated before. The most relevant works are recent theoretical investigations \cite{coherent control of SPP} on the coherent control of the group velocity and related phenomena of surface polaritons at the interface of an EITM/negative refractive index metamaterial, in which case there is extremely low loss of the polaritons at a certain frequency due to the double negative index of the metamaterial (i.e. $\mu<0$ and $\varepsilon<0$). For an ordinary metal, however, because only $\varepsilon$ can be negative, there is no such interference effect to eliminate losses in the metal. Hence a question arises: for SP polaritons at a metal surface, will coherent control based on EIT be possible, or will some other coherent phenomena be observed? It will be shown in this letter that coherent control is indeed tolerant of metal loss, and some very interesting phenomena may be observed in an EIT-SPR system.

A schematic diagram of the EIT-SPR system is shown in Fig.1. We consider the excitation of SP polaritons in the Kretschmann geometry of the attenuated total reflection  method \cite{Raether}. The system is composed of three layers: a prism, a thin metal film (50 nm thick), and a dielectric. The last is assumed to be a hybrid medium which is composed of three-level atoms and a background material of permittivity $\varepsilon_{\rm b}$. Two monochromatic laser beams, a probe and a coupling beam propagating through the prism are both incident on the metal film and then coupled to the atoms (see Fig.1(a)). Each beam can excite a surface plasmon polariton (SPP) under resonance conditions.

We first consider a weak probe field polariton [transverse magnetic (TM) mode] at the interface between a homogeneous dielectric and a metal of semi-infinite medium, as shown in Fig. 1(b). From the surface boundary conditions for an SPP wave vector, its magnetic field can be given by \cite{EM SPP,Nonlinear SPP}
\begin{equation}
H_y=H_{y0}e^{ik_{\rm spp}z \pm\kappa_{\rm m,d}x}
\label{Hy}
\end{equation}
where '$+$' is for $x<0$ and '$-$' is for $x>0$, $H_{y0}$ is field-amplitude,
\begin{equation}
k_{\rm spp}=k_0\sqrt{\f{\varepsilon_{\rm d} \varepsilon_{\rm m}}{\varepsilon_{\rm d}+\varepsilon_{\rm m}}};
\kappa_{\rm m,d}=\sqrt{k_{\rm spp}^2-k_0^2\varepsilon_{\rm m,d}}
\label{kappa}
,\end{equation}
 where $k_0=\f{\omega_p}{c}$, $\omega_p$ is the angular frequency of the probe field, $\varepsilon_{\rm d}$ and $\varepsilon_{\rm m}$ are the permittivities of the dielectric and the metal, respectively. For the EIT case, $\varepsilon_{\rm d}$ will be coherently controlled by the coupling field.
 Because  the EIT `transparency' range is  very narrow (about 100 MHz (see Fig. 2)), the frequency dependence of $\varepsilon_{\rm m}$ can be safely neglected.
The probe field wavelength is taken to be about 589 nm, where, according to experimental data \cite{handbook}, $\varepsilon_{\rm m}=-13.3 + 0.883i$, and the imaginary part accounts for metal loss. It will be found later that the coupling field can be another SP polariton or a freely-propagating traveling field, but only in the former case will $\varepsilon_{\rm d}$  be spatially independent and
Eq. (\ref{kappa}) valid.

The EITM can be an atomic gas or doped
solid medium composed of three-level atoms of $\Lambda$ configuration.
According to the Schr$\ddot{o}$dinger equation in the interaction picture, the motion of amplitudes $c_0$, $c_1$, and $c_2$ can be given by \cite{EITreview}
\begin{equation}
\begin{array}{lcr}
i \f{dc_0}{dt}=\f{\Omega_p}{2}c_1\\
i \f{dc_1}{dt}=-(\delta_p+i\f{\gamma}{2})c_1+\f{\Omega_p}{2}c_0+\f{\Omega_c}{2}c_2\\
i\f{dc_2}{dt}=-(\delta_p-\delta_c+i\f{\gamma'}{2})c_2+\f{\Omega_c}{2}c_1
\end{array} ,
\label{amplitude}
\end{equation}
Here $\delta_i$ are detunings defined by $\delta_p=\omega_p-\omega_{10}$ and $\delta_c=\omega_c-\omega_{12}$, where $\omega_p$ and $\omega_c$ are the angular frequencies of the probe and coupling fields, respectively, $\omega_{10}$ and $\omega_{1,2}$ are the angular atomic transition frequencies, $\Omega_i$ is the Rabi frequency defined by $\Omega_p=\f{e{\bf x}_{10}\cdot {\bf E}_p}{2\hbar}$,$\Omega_c=\f{e{\bf x}_{21}\cdot {\bf E}_c}{2\hbar}$, and ${\bf E}_i$ is the electric field amplitude with $i=p,c$ denoting the probe and coupling fields, respectively.  Also, $\gamma$ and $\gamma'$ are the decay rates of  states $|1>$ and $|2>$, respectively. The quantum interference between $|0>-|1>$ and $|1>-|2>$ transitions can strongly modify the optical response of the system. For the sake of simplicity, we assume the EITM is composed of cold atoms where the Doppler effect can be neglected, and only consider the weak probe field case where $\Omega_p<<\gamma, \Omega_c$.
Here we only consider very dilute atomic medium as in Ref. \cite{slow}.  The permittivity of the atomic medium can be given by
\begin{equation}
\varepsilon_{\rm d}=\varepsilon_{\rm b}+N\alpha
\label{eps}
\end{equation}
where $\varepsilon_{\rm b}$ ($\simeq 1$) is the background permittivity (of vacuum or a dilute background dielectric) and
\begin{equation}
\alpha=-\f{\mu^2 }{\hbar\varepsilon_0} \f{c_0^*c_1}{\Omega_p}=
-\f{\mu^2 }{\hbar\varepsilon_0}\f{\f{1}{2}(\delta_p-\delta_c+i\gamma'/2)}
{\[-1/4\Omega_c^2+(\delta_p+i\gamma/2)(\delta_p-\delta_c+i\gamma'/2)\]};
\label{alpha}
\end{equation}

Here $N$ is the atomic number density, and
$\mu=ex_{10}= \sqrt{\f{1}{3}}3.5247ea_0$ \cite{Steck} for transitions of the sodium D2 line \cite{Steck}, with $a_0$ being the Bohr radius and $\varepsilon_0$ the
permittivity of vacuum.
For the D2 line of sodium, from the experimental data of Ref. \cite{Steck}, $\gamma=61.54$MHz, and from that of \cite{slow}, $N\simeq 3\times 10^{18}/m^3$ .
If $\delta_p\simeq\delta_c$, then Im$(\varepsilon)\simeq 0$ but Re$(\varepsilon)$ varies rapidly with $\delta_p$ or $\delta_c$, which leads to strong dispersion within the transparency window.
        It should be noted that, because the atoms have been assumed to be prepared initially in the ground state $|0>$ and $\Omega_p<<\Omega_c$, the atoms approach steady state rapidly and the population in state $|2>$ is very small at all times.  Also, the atomic medium is dilute but the coupling field is strong, so the influence of the former on the latter can be safely neglected.

It should be emphasized that the amplitude of the electric component of the coupling field ${\bf E}_c$ is spatially independent for freely propagating (traveling) fields, but decays exponentially with $x$ for SP polaritons. Only in the former case is $\varepsilon_{\rm d}$ spatially independent and Eq. (\ref{kappa}) valid.   For a strong coupling-field SPP (where the dielectric acts as a vacuum), $\Omega_c(x)=\Omega_{c0}e^{{\rm Re}[\kappa_c^{(0)}] x}$, then $\varepsilon_{\rm d}$ will depend on $x$.
If the variation of the refractivity satisfies $\delta n_{\rm d} <<n_{\rm b} \simeq 1$, then the SPP propagation constant $k_{\rm spp}$ can be calculated by perturbation theory \cite{perturbation}:
\begin{equation}
k_{\rm spp}=k_{\rm spp}^{(0)}+\delta k_{\rm spp}
\label{deltakspp}
\end{equation}
where
\begin{equation}
\delta k_{\rm spp}=\f{1}{2}\f{k_0^2}{k_{\rm spp}^{(0)}}{\rm Re}\[\kappa_p^{(0)}\]\int_{-\infty}^0e^{2{\rm Re}\[\kappa_p^{(0)}\] x}\[\varepsilon_{\rm d}(x)-1\]dx.
\label{deltakspp}
\end{equation}
According to Eq. (\ref{kappa}) the zeroth-order propagation constant and confinement are
 $k_{\rm spp}^{(0)}(\omega_i)=\f{\omega_i}{c}\sqrt{\f{\varepsilon_{\rm m}(\omega_i)}{1+\varepsilon_{\rm m}(\omega_i)}}$, and  $\kappa_{i}^{(0)}=\f{\omega_{i}}{c}\sqrt{\f{-1}{1+\varepsilon_{\rm m}(\omega_{i})}}$ , respectively ($i=p,c$ denote `probe' and `coupling' ,respectively).

If the difference of the two ground states is very small compared with the laser frequencies, e.g. 1.8GHz in the experiment of Ref.\cite{slow}, then $\kappa_p^{(0)} \approx \kappa_c^{(0)} $, where the superscript $0$ denotes `zeroth order', i.e. the case when the dielectric is vacuum.
Thus we  obtain
 \begin{equation}
  \delta k_{\rm spp}= \f{1}{2}\f{k_0^2}{k_{\rm spp}^{(0)}}\[\varepsilon_{\rm b}-1+\f{\mu^2N}{\hbar \varepsilon_0}\f{\beta}{\delta_p+i\gamma/2}{\rm ln}(1-1/\beta)\]
\label{deltakspp2}
 \end{equation}
where
\begin{equation}
\beta=\f{4(\delta_p+i\gamma/2)(\delta_p-\delta_c+i \gamma'/2)}{\Omega_c^2}
\label{beta}
\end{equation}

We now calculate the reflectivity $R$ of the probe beam defined by $R=|r_{\rm pmd}|^2 $, where $r_{\rm pmd}$ is the three-layer amplitude reflection coefficient and is given by the Fresnel formula
\begin{equation}
r_{\rm pmd}=\f{r_{\rm pm}+r_{\rm md}{\rm exp}(2ik_{{\rm m}x}q)}{1+r_{\rm pm}r_{\rm md}{\rm exp}(2ik_{{\rm m}x}q)}
\label{rpmd}
\end{equation}
where $q$ is the thickness of the metal film, and the two-layer amplitude reflection coefficients $r_{\rm pm}$ and $r_{\rm md}$ at the prism/metal and metal/dielectric interfaces, respectively, are given by\begin{equation}
r_{\rm pm}=\f{\varepsilon_{\rm m} k_{px}-\varepsilon_{\rm p} k_{{\rm m}x}}{\varepsilon_{\rm m} k_{px}+\varepsilon_{\rm p} k_{{\rm m}x}};
r_{\rm md}=\f{\varepsilon_{\rm d} k_{{\rm m}x}-\varepsilon_{\rm m} k_{{\rm d}x}}{\varepsilon_{\rm d} k_{{\rm m}x}+\varepsilon_{\rm m} k_{{\rm d}x}}.
\label{rij}.
\end{equation}
Here $k_z$ and $k_{jx}$ are the parallel and normal wave vectors which are given by
\begin{equation}
k_z=k_0n_{\rm p} {\rm sin}\theta_p; k_{jx}=\sqrt{k_0^2\varepsilon_j-k_z^2} \label{kjx}
\end{equation}
with $j={\rm p, m, d}$ denoting prism, metal, and dielectric (EIT medium), respectively. The field enhancement due to surface plasmon is defined by \cite{Raether}
\begin{equation}
T=\l|\f{H_y(m/d)}{H_y(p/m)}\r|^2=\f{t_{\rm pm}t_{\rm md}{\rm exp}(ik_{{\rm m}x}q))}{1+r_{\rm pm}r_{\rm md}{\rm exp}(2ik_{{\rm m}x}q)},
\label{enhancement}.
\end{equation}
where $t_{ij}=1+r_{ij}$which is derived from the boundary conditions,  and $i,j={\rm p,m}$ or $i,j={\rm m,d}$.
The enhancement of the electric field can be given by $T_e=\f{\varepsilon_{\rm p}}{\varepsilon_{\rm d}}T$. The reflectivity of the probe laser is strongly influenced by the coupling laser via the quantum interference effect in the EIT medium. The coupling field in the dielectric can be a freely propagating field or an evanescent field, depending on the incident angle $\theta_c$. The critical case occurs when $k_{{\rm d}x}^2=0$, i.e. $n_{\rm p} \sin\theta=\sqrt{\varepsilon_{\rm d}}$, which gives the critical angle $\theta_{crt}$. For $n_{\rm p}=1.51$ and $\varepsilon_{\rm m}=-13.3 + 0.883i$,  $\theta_{crt}=41.47^{\rm o}$.
The coupling field will be an evanescent field if $\theta_c>\theta_c^{(crt)}$, but will be a freely propagating field if $\theta_c<\theta_c^{(crt)}$.
 For $\theta_c=0$ the field enhancement $T_e \approx 0.043$ and $0.21 $ for 50nm and 30nm silver films,respectively. For $\theta_c\rightarrow \theta_c^{({crt})}$, $T_e\approx 0.47 $ and $ 2.2$ for silver films of thicknesses 50 nm and 30 nm, respectively. In practice, the EIT-SPR system may be realized with three different schemes using different arrangements of the coupling field, as follows. (1) With thin metal films, a TE-mode coupling field and a TM-mode probe field, which allows for easy separation of the reflected probe and coupling beams with a polarizing beam-splitter.
 (2) With thicker metal films (e.g. 50 nm thick Ag ) and a TM ($p$-polarized) coupling field with the incidence angle $\theta_c\approx\theta_c^{(crt)}$, in which case the large field enhancement factor $T_e$ ensures a strong coupling field.
  (3) With an incidence angle of $\theta_c\approx \theta_c^{res}$, where $\theta_c^{res}$ is the resonance angle for excitation of SP polaritons which can be evaluated by the dip in the angle spectrum of the reflectivity $R$. The other conditions are the same as in scheme (2). One of the advantage of scheme (3) is  the very large field enhancement factor $T_e$ (in resonance case $T_e\sim 100$ for 50 nm silver film), which ensures that a large $\Omega_c$ can be easily obtained, even when the coupling beam is very weak, which is of interest for quantum information processing.
In schemes (1) and (3), $\Omega_c$ is constant, so the propagation constant $k_{\rm spp}$ and the reflectivity $R$ may be easily obtained  using Eqs. (\ref{kappa}), (\ref{rpmd}), (\ref{rij}),(\ref{kjx}), (\ref{eps}), and (\ref{alpha})(see blue curves in Fig.2(b)and(c)).

Only in scheme (3) will $\Omega_c$ be spatially dependent. In this case, to calculate $r_{\rm md}$ we consider
a probe field of  $H_{Iy}(x,z)=A_Ie^{-i{k_{{\rm v}x}} x-ik_z z}$ which is reflected from the vacuum/dielectric interface(here `v' denotes `vacuum'). If $\varepsilon_{\rm d}(x)\simeq 1$,  with perturbation theory similar to that in Ref. \cite{perturbation}, the reflection field can be written as
 \begin{equation} H_{Ry}(x,z)=A_Ie^{ik_z z}k_0^2\int_{-\infty}^0G(x-x')\[\varepsilon_{\rm d}(x')-1\]e^{-ik_{{\rm v}x} x'}dx'
\label{Hy}
\end{equation}
 where $A_R^{(0)}$ is the zeroth-order amplitude of the reflected field, i.e. that in the absence of an EIT medium .The Green function $G(x-x')$ is given by
\begin{equation}
G(x-x')=\f{i}{2k_{{\rm v}x}}e^{ik_{{\rm v}x} (x-x')}
\label{G}
\end{equation}
Substituting Eq. (\ref{G}) into (\ref{Hy}), we obtain $r_{{\rm vd}}$
\begin{equation}
r_{{\rm vd}}=\f{H_{Ry}(0,z)}{H_{Iy}(0,z)}=\f{k_0^2}{k_{{\rm v}x}^2}\[\varepsilon_{\rm b}-1-\f{\mu^2N}{\hbar \varepsilon_0}\f{1}{\delta_p+i\gamma/2}F(1,b;1+b;1/\beta)\]
\label{rmdg}
\end{equation}
where $F$ is the general hypergeometric function\cite{int},
$b=-i{k_{{\rm v}x}}/{\rm Im}[{k_{{\rm v}x}}^{(\rm c)}]$,
$ k_{{\rm v}x}=\sqrt{k_0^2-k_z^2}$ ,and $ {k_{{\rm v}x}}^{\rm (c)}=\sqrt{\f{\omega_c^2}{c^2}-{k_z^{\rm (c)}}^2}$.
 In the typical resonance cases when  ${k_{{\rm v}x}}\simeq {k_{{\rm v}x}}^{(\rm c)}\simeq i\kappa_p^{(0)} $,
 from Eqs. (\ref{G} ),(\ref{Hy}),and (\ref{rmdg}), it is easy to see that
\begin{equation}
r_{{\rm vd}}=\f{k_0^2}{{k_{{\rm v}x}}^2}\[\varepsilon_{\rm b}-1-\f{\mu^2N}{\hbar \varepsilon_0}\f{\beta}{\delta_p+i\gamma/2}{\rm ln}(1-1/\beta)\]
\label{rvdn}
\end{equation}
where the superscript $(0)$ denotes zero order, i.e. the case that EIT atoms are absent, and $\beta$ is given by Eq. (\ref{beta}).
Similar  to Eq. (\ref{rpmd}), the three-layer (metal/vacuum/dielectric) amplitude reflectivity  is
 $r_{\rm mvd}=\f{r_{\rm mv}+r_{\rm vd}{\rm exp}(2ik_{\rm vx}q)}{1+r_{\rm mv}r_{\rm vd}{\rm exp}(2ik_{\rm vx}q)}$, where `v' denotes the vacuum layer, $q$ is its thickness.
Setting $q=0$ we obtain two-layer (metal/dielectric) amplitude reflectivity $r_{\rm md}$ :
\begin{equation}
r_{\rm md}=\f{r_{\rm mv}+r_{{\rm vd}}}{1+r_{\rm mv}r_{{\rm vd}}}
\label{rmdn}
,\end{equation}
where
$ r_{\rm mv}=\f{k_{{\rm m}x}-\varepsilon_{\rm m} k_{{\rm v}x}}{k_{{\rm m}x}+\varepsilon_{\rm m} k_{{\rm v}x}}$.
Substituting Eq. (\ref{rmdn}) into Eq. (\ref{rpmd}) and repeating the foregoing procedure, we can complete all the calculations for the propagation constant $k_{\rm spp}$ of the probe-field SP polaritons and the reflectivity $R$ of the probe beam. Their dependence on the probe detuning $\delta_p$ are shown in Fig. 2.
 Within the EIT transparency window, Im$ (k_{\rm spp})\simeq 0$, i.e. the polaritons only suffer low losses. If the coupling field is a freely propagating field (blue curves), the profile of $k_{\rm spp}(\delta_p)$ (b) is very similar to that of the refractivity $n(\delta_p)$(a) of bulk EITM. The ultranarrow linewidth of the EIT window leads to a very sharp dip in the reflectivity spectrum $R(\delta_p)$ at $\delta_p=0$,  where $R$ is extremely sensitive to the variation of $\delta_p$.

 It is interesting that if the coupling field is an SPP instead, the gradient of the Re$\[k_{\rm spp}(\delta_p)\]$ curve will be steeper, and the dips in Im[$k_{\rm spp}(\delta_p)]$ and the $R(\delta_p)$-curves will be more pointed (see red curves in Figs. 2 (b) and (c)).  The latter arise from the strong confinement of the coupling field and are absent in ordinary EIT systems. They are signatures of the two-dimensional EIT and should be observable experimentally.

 If both the lasers are monochromatic, then the variations of $\delta_p$ and $\delta_c$  account for atomic level shifts induced by environmental fields, e.g. by DC magnetic fields via the Zeeman effect.
 It should be emphasized that a probe-field SPP strongly confined in the metal/EITM interface only responds to a DC magnetic field very near the interface (within a distance of about $1/{\rm Re}\[\kappa_p^{(0)}\]$. Consequently, it may be possible to apply this EIT-SPR system in novel magnetometers for highly localized measurements.

The propagation constant Re$(k_{\rm spp})$  and thus the reflectivity spectrum $R$ are also extremely sensitive to variations in the substrate permittivity $\varepsilon_{\rm b}$.
For a dilute gas EITM, a small increase in $\varepsilon_{\rm b}$ can be caused by another background dilute gas mixed with the EIT gas. It is shown that a variation of only 5/1000 of $\epsilon_{\rm b}$ can dramatically  change the reflectivity spectrum (see Fig.2(c), red curve for $\varepsilon_{\rm b}=0$, while black curve for $\varepsilon_{\rm b}=1.005$). Although $\varepsilon_{\rm b}$  depends on $\omega_p$ in general,  it is frequency independent within the ultra-narrow EIT transparency window.   The substrate sensitivity of the spectrum may possibly be used for chemical or biological sensors.

The angle dependency of the reflectivity $R(\theta_p)$ is shown in Fig. 3.
When the dielectric is an EITM and  $\delta_p=0$, the $R(\theta_p)$ curve is very close to that when there is a vacuum. If the probe field is slightly detuned from resonance, e.g. $\delta=0.5\gamma$, the resonance angle (i.e. the dip in the $R(\theta_p)$ curve) will be significantly shifted . For the two-level case, however, the minimum of $R$ is much larger than in the EIT case.

The detuning spectrum of the reflectivity $R$ is also strongly dependent on the atomic number density of the EITM , as shown in Fig. 4.  In the two-level case there is a peak in the $R(\delta_p)$ spectrum, which becomes broader with the increase of the atomic number density $N$. In the EIT case, however,  a very narrow dip appears at the center of the background peak. It is interesting that as $N$ increases the dip becomes narrower  but the background peak becomes broader.

In summary, an EIT-SPR system has been analytically investigated. The  quantum interference between the two transition paths of the EIT atoms can strongly modify the properties of the surface plasmon polaritons at the metal/EITM interface, and lead to new interesting properties of the reflectivity spectrum. The probe field SPP and its reflectivity are extremely sensitive to shifts in laser frequency or atomic levels, and to variations of the permittivity of the background constituents mixed  with the EIT medium. The profile of the reflectivity frequency spectrum is different for freely propagating and SPP coupling fields.  All parameters used in this letter were taken from experimental data in published works and losses in the metal have been considered. The results indicate a route to the realization of low-dimensional on-chip EIT systems,  and possible future applications in extremely sensitive magnetometers and biosensors with highly localized responses.

This work is partially supported by the National Natural Science Foundation of
China (Grant No. 10504016) and the ¡°973¡± program (Grant No.
2010CB922904) of the Ministry of Science and Technology
(MOST), China.

\newpage
\baselineskip 24pt \normalsize
\begin{figure}[htpb] \begin{center}
\includegraphics[width=14cm]{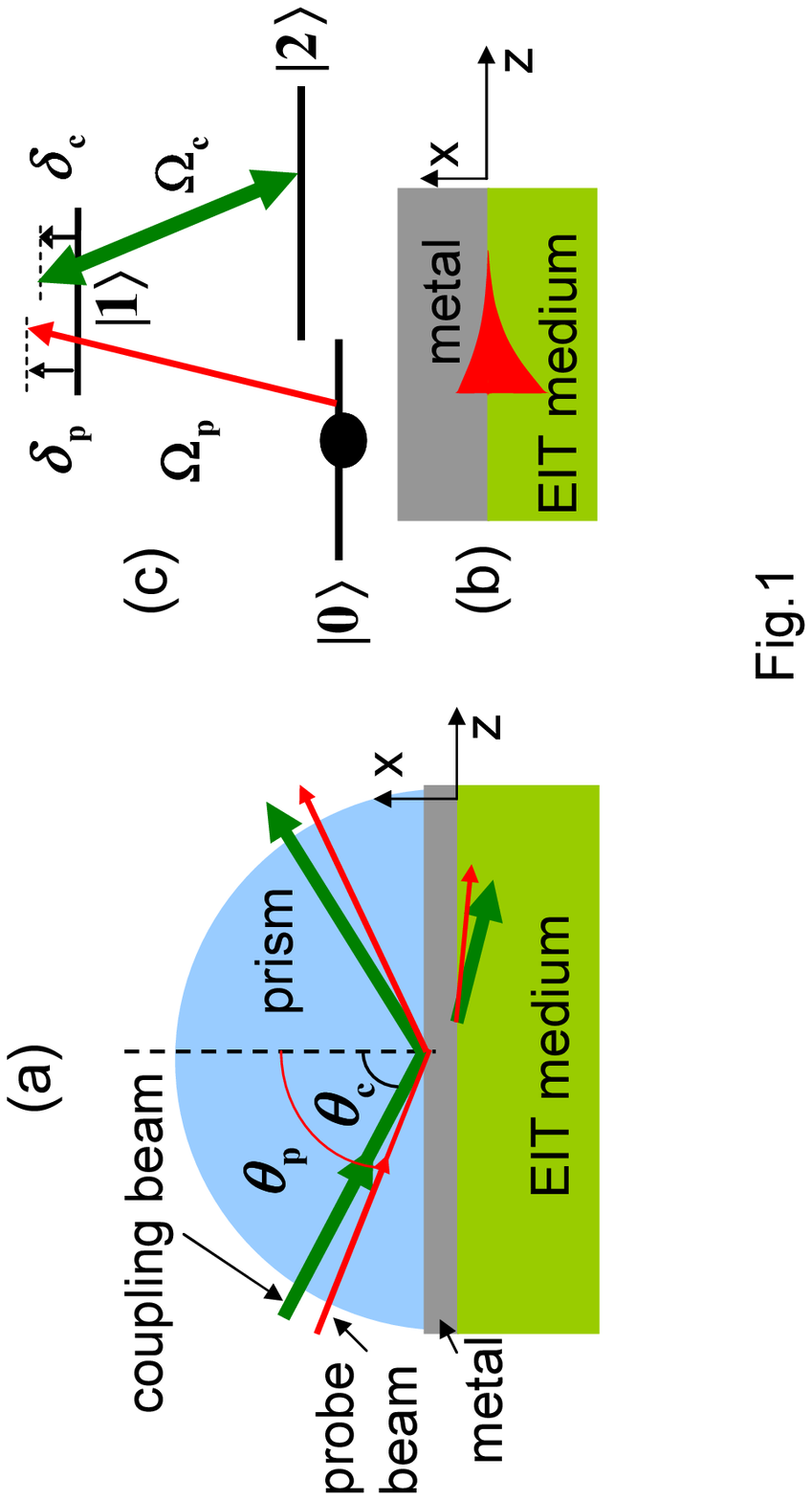}
\caption{(Color online) Schematic of an EIT-SPR system. (a) SPR under EIT conditions, where the probe and coupling laser beams both couple to the EIT atoms. The blue area is a cylindrical glass prism. (b) Weak SPP (TM mode) probe field propagating along and confined at the interface between a metal and a  semi-infinite EIT medium. (c) Energy-level diagram of three-level atoms of $\Lambda$-configuration.} \label{Fig.1} \end{center} \end{figure}

\begin{figure}[htpb] \begin{center}
\includegraphics[width=18cm]{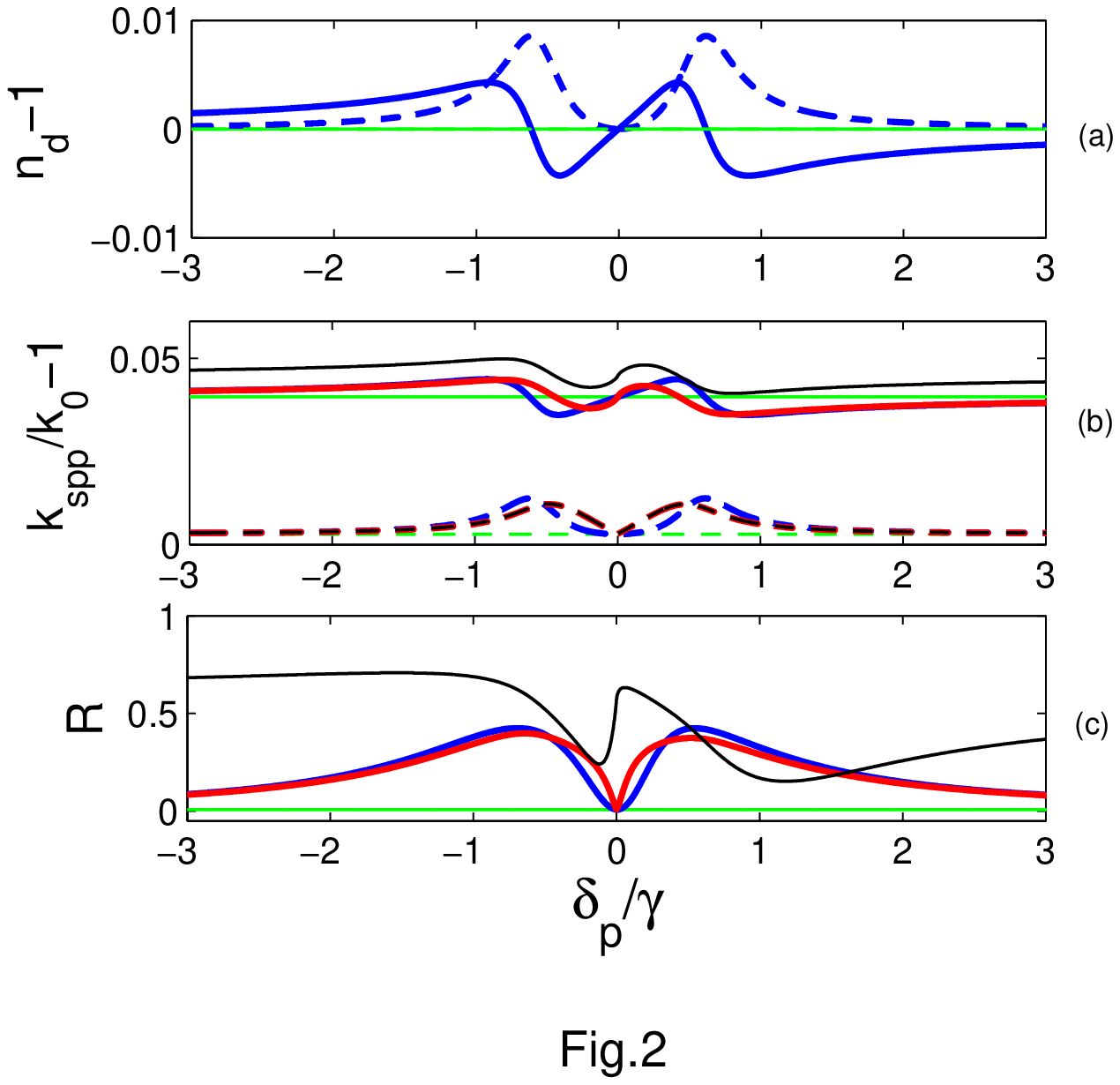}
\caption{(Color online) Schematic of an EIT-SPR system. (a) SPR under EIT conditions, where the probe and coupling laser beams both couple to the EIT atoms. The blue area is a cylindrical glass prism. (b) Weak SPP (TM mode) probe field propagating along and confined at the interface between a metal and a  semi-infinite EIT medium. (c) Energy-level diagram of three-level atoms of $\Lambda$-configuration.} \label{Fig.2} \end{center} \end{figure}

\begin{figure}[htpb] \begin{center}
\includegraphics[width=20cm]{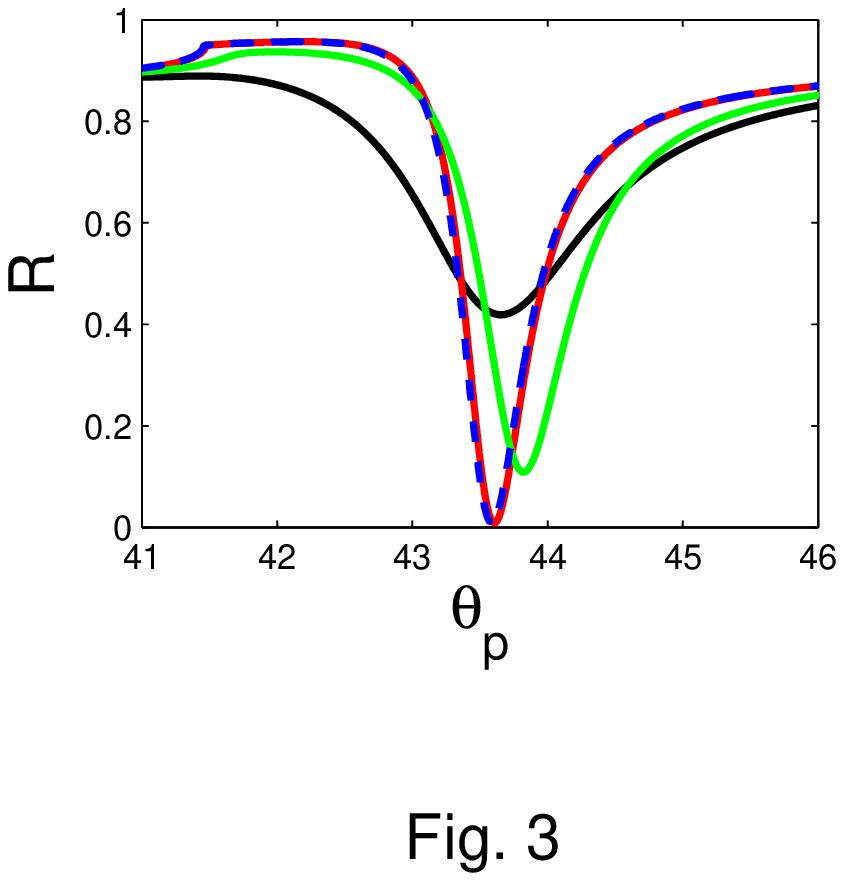}
\caption{(Color online) Angle-dependence of the reflectivity $R$. Blue dashed curve: for a vacuum dielectric; red and green solid curves: for an EIT medium with $\delta_p=0$, and $0.3\gamma$, respectively; black curve: for a two-level atomic medium with $\delta_p=0$. Other parameters are the same as in Fig.2.} \label{Fig.3} \end{center} \end{figure}

\begin{figure}[htpb] \begin{center}
\includegraphics[width=14cm]{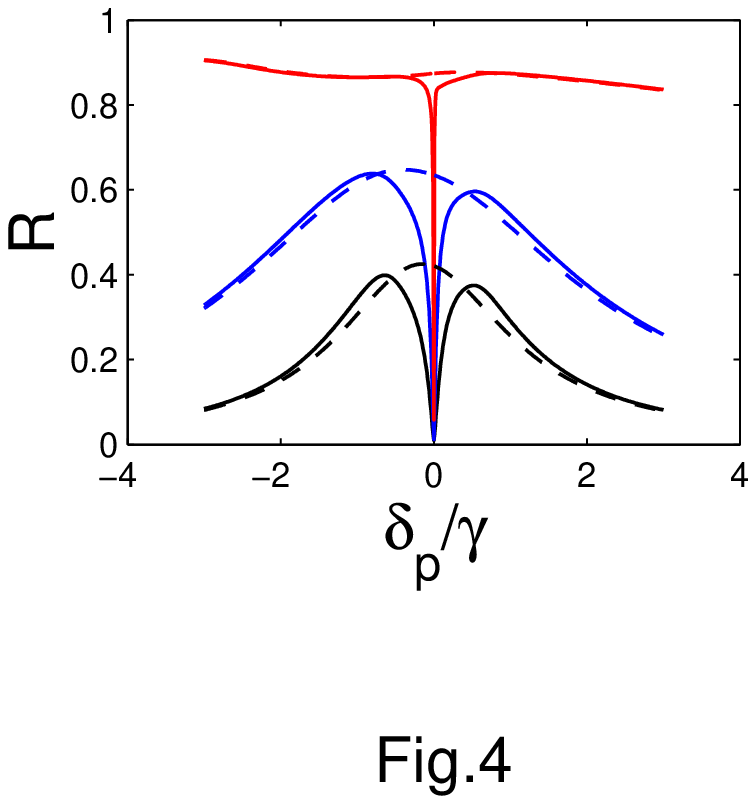}
\caption{(Color online) Angle-dependence of the reflectivity $R$. Blue dashed curve: for a vacuum dielectric; red and green solid curves: for an EIT medium with $\delta_p=0$, and $0.3\gamma$, respectively; black curve: for a two-level atomic medium with $\delta_p=0$. Other parameters are the same as in Fig.2} \label{Fig.4} \end{center} \end{figure}
\end{document}